\documentclass{ws-procs975x65}

\def\beq{\begin{equation}}
\def\eeq{\end{equation}}

\begin{document}

\def \d {{\rm d}}
\def \q {Q}
\def \etta {F} 
\def \t {{\Theta}}
\def \k {{\kappa}}
\def \l {{\lambda}}
\def \s {{\sigma}}

\def \tv {{\tilde v}}
\def \tz {{\tilde z}}

\def \P {{p_\lambda}}
\def \Q {{p_\nu}}
\def \eigen {{\mathcal N}}
\def \M {{\mathcal M}}
\def \bm #1 {\mbox{\boldmath{$m^{(#1)}$}}}

\def \bF {\mbox{\boldmath{$F$}}}
\def \bk {\mbox{\boldmath{$k$}}}
\def \bl {\mbox{\boldmath{$l$}}}
\def \bbm {\mbox{\boldmath{$m$}}}
\def \tbbm {\mbox{\boldmath{$\bar m$}}}

\newcommand{\be}{\begin{equation}}
\newcommand{\ee}{\end{equation}}

\newcommand{\beqn}{\begin{eqnarray}}
\newcommand{\eeqn}{\end{eqnarray}}
\newcommand{\AdS}{anti--de~Sitter }
\newcommand{\AAdS}{\mbox{(anti--)}de~Sitter }
\newcommand{\AAN}{\mbox{(anti--)}Nariai }
\newcommand{\AS}{Aichelburg-Sexl }
\newcommand{\pa}{\partial}
\newcommand{\pp}{{\it pp\,}-}
\newcommand{\ba}{\begin{array}}
\newcommand{\ea}{\end{array}}

\newcommand{\tr}{\textcolor{red}}
\newcommand{\tb}{\textcolor{blue}}
\newcommand{\tg}{\textcolor{green}}

\def\a{\alpha}
\def\g{\gamma}
\def\de{\delta}

\def\b{{\kappa_0}}

\def\E{{\cal E}}
\def\B{{\cal B}}
\def\R{{\cal R}}
\def\F{{\cal F}}

\def\e{e}
\def\bb{b}

\title{Higher dimensional Robinson-Trautman spacetimes sourced by $p$-forms: static and radiating black holes}

\author{Marcello Ortaggio} 
\address{Institute of Mathematics of the Czech Academy of Sciences \\ \v Zitn\' a 25, 115 67 Prague 1, Czech Republic \\
$^*$E-mail: ortaggio(at)math(dot)cas(dot)cz}

\author{Ji\v{r}\'{\i} Podolsk\'y and Martin \v{Z}ofka}
\address{Institute of Theoretical Physics, Faculty of Mathematics and Physics, \\
Charles University in Prague, V Hole\v{s}ovi\v{c}k\'{a}ch 2, 180 00 Prague 8,  Czech Republic \\
$^*$E-mail: podolsky(at)mbox(dot)troja(dot)mff(dot)cuni(dot)cz, zofka(at)mbox(dot)troja(dot)mff(dot)cuni(dot)cz}

\begin{abstract}
We summarize results about Robinson--Trautman spacetimes in the presence of an aligned $p$-form Maxwell field and an arbitrary cosmological constant in $n\ge 4$ dimensions. While in {\em odd} dimensions the solutions reduce to static black holes dressed with an electric and a magnetic field (with an Einstein space horizon), in {\em even} dimensions $2p=n$ they may also describe black holes gaining (or losing) mass by receiving (or emitting) electromagnetic radiation. The Weyl type of the spacetimes is also briefly discussed in all the possible cases. 
\end{abstract}

\bodymatter

\section{Introduction}

The Einstein-Maxwell theory can be generalized by considering $p$-form fields $\bF$ coupled to gravity for an arbitrary $0<p<n$ ($n$ is the number of spacetime dimensions), cf., e.g., \cite{Teitelboim86}. This is of particular interest when $n>4$. The corresponding Maxwell equations, in the source-free case (to which we will restrict hereafter), read
\be
  (\sqrt{-g}\,F^{\mu\a_1\ldots\a_{p-1}})_{,\mu}=0 , \qquad F_{[\a_1\ldots\a_p,\mu]}=0 .
	\label{Maxwell}
\ee 
The back-reaction of $\bF$ enters Einstein's equations as
\begin{equation}\label{Ricci}
	R_{\mu\nu} = \frac{2}{n-2} \Lambda g_{\mu\nu} + \b\left[F_{\mu\a_1\ldots\a_{p-1}} {F_\nu}^{\a_1\ldots\a_{p-1}}-\frac{p-1}{p(n-2)} g_{\mu\nu}F^2 \right] ,
\end{equation}
where $F^2=F_{\a_1\ldots\a_{p}}F^{\a_1\ldots\a_{p}}$, $\b$ is a coupling constant and $\Lambda$ is the cosmological constant.

Electric static black holes in the standard case $p=2$ were studied already in the early paper \cite{Tangherlini63}, and a magnetic ``charge'' (in even dimensions) was added in \cite{OrtPodZof08}. More recently, it has been shown that asymptotically flat static black holes cannot couple to electric $p$-form fields when $(n+1)/2\le p\le n-1$ (and thus do not possess dipole hair) \cite{EmpOhaShi10} and that, for any $p>2$, static perturbations of the vacuum Schwarzschild--Tangherlini metric do not exist \cite{Guven89,ShiOhaTan11}. Subsequently, a large family of $p$-form black holes (with various possible horizon geometries and asymptotia) has been constructed in \cite{BarCalCha12} (see \cite{OrtPodZof08} for related earlier work in the case $p=2$). In addition, results of \cite{Durkeeetal10} showed that electromagnetic radiation may have properties different from those of standard $n=4$, $p=2$ electrovac general relativity (except for $2p=n$), as confirmed in \cite{Ortaggio14} in the case of test fields.

Our recent work \cite{OrtPodZof15}, to be summarized here, considered exact solutions of $p$-form Einstein-Maxwell gravity in the broader context of Robinson-Trautman spacetimes, which allow also for radiative (non-static) solutions. The Robinson-Trautman class  is defined by the existence of a geodesic, shear-free, twist-free but expanding null vector field $\bk$ and was first studied in $n=4$ general relativity \cite{RobTra62} (see also the reviews in \cite{Stephanibook,GriPodbook}). This was extended to an arbitrary $n$ in\cite{PodOrt06} and further studied in \cite{Ortaggio07,OrtPodZof08,PraPraOrt07,SvaPod14,PodSva15}. Some of the general results of \cite{PodOrt06,PodSva15} have been summarized in Theorem~1 of Appendix~A of \cite{OrtPodZof15}, which may be useful to recall here:

\begin{theorem}[Robinson-Trautman spacetimes of aligned Ricci type II]
\label{theor_RT}

If a $n$-dimensional spacetime {($n\ge 4$)} admits a non-twisting, non-shearing, expanding geodesic null vector field~{\boldmath $k$} and the Ricci tensor is of aligned type II, adapted coordinates $(u,r,x^1,\ldots,x^{n-2})$ can be chosen such that \cite{PodOrt06}
\beqn
  & & \d s^2=r^{2}h_{ij}\left(\d x^i+ W^{i}\d u\right)\left(\d x^j+ W^{j}\d u\right){-}2\,\d u\d r-2H\d u^2 , \label{geo_metric} \\
	& & h_{ij}=h_{ij}(u,x) , \qquad W^{i}={\alpha}^i(u,x)+r^{1-n}{\beta}^i(u,x) , \label{h_W} \\
	& & \bk=\pa_r , \qquad \theta=1/r ,
\eeqn
where $H$ is an arbitrary function of all coordinates. $\bk$ is automatically a WAND, such that the Weyl tensor is in general of aligned type I(b). It is a multiple WAND iff ${\beta}^i=0$ \cite{PodSva15}, in which case the Weyl tensor is of aligned type II(d) (or more special). {When ${\beta}^i=0$, one can locally set $W^{i}=0$ (after a coordinate transformation giving ${\alpha}^i=0$) \cite{PodOrt06}.} {The Weyl type further specializes to II(bd) iff $h_{ij}$ is an Einstein metric (with still $W^{i}=0$) \cite{PodSva15}.}

\end{theorem}

Here and in the following, the vector field {\boldmath $k$} is the generator of null hypersurfaces $u=\,$const such that $k_\mu\d x^\mu={-}\d u$, $r$ is an affine parameter along $\bk$, $\theta$ is its expansion scalar, and $x \equiv (x^i) \equiv (x^1, \ldots, x^{n-2})$ are spatial coordinates on a ``transverse'' $(n-2)$-dimensional Riemannian manifold.  For certain calculations it may be useful to observe that $2H=g^{rr}=-g_{uu}$ and $W^i=g^{ri}$ (such that $W^i=0\Leftrightarrow g^{ri}=0\Leftrightarrow g_{ui}=0$). Recall that in {\em vacuum} or with {\em aligned pure radiation} necessarily ${\beta}^i=0$ {and $h_{ij}$ is Einstein} for any $n\ge 4$, and the Weyl tensor further specializes to type D(bd) {(possibly, D(bcd) or D(acd))} if $n>4$, cf.~\cite{PodOrt06}. If one relaxes the assumptions of the theorem by requiring only the aligned Ricci type I (i.e., $R_{rr}=0$), one obtains the same form of the metric, except that the $W^i(u,r,x)$ are arbitrary functions  \cite{PodOrt06}.

The above theorem includes, in particular, the case of Robinson-Trautman spacetimes in the presence of an aligned $p$-form $\bF$, i.e., such that 
\be
	F_{r i_1\ldots i_{p-1}}=0 . 
	\label{F_aligned}
\ee
The corresponding Einstein-Maxwell equations have been studied in\cite{OrtPodZof15}. As it turns out, one has to consider separately the case of a generic $n$ (section~\ref{sec_summary_gen}), and the special case of an even $n=2p$ (section~\ref{subsec_2p=n_summary}). The case $p=1$ is special in all dimensions and leads to static metrics with a non-Einstein transverse space, see\cite{OrtPodZof15} for details.

\section{Generic case $2p\neq n$ ($n>4$): static black holes}

\label{sec_summary_gen}

The line-element is given by 
\be
  \d s^2=r^{2}h_{ij}\d x^i\d x^j{-}2\,\d u\d r-2H\d u^2 , \label{ds_generic}
\ee
where $h_{ij}=h_{ij}(x)$ is the metric of a Riemannian Einstein space of dimension $(n-2$) and scalar curvature $\R=K(n-2)(n-3)$, and 
\begin{eqnarray}
	& & 2H=K-\frac{2\Lambda}{(n-1)(n-2)}\,r^2-\frac{\mu}{r^{n-3}} \nonumber\\
	& & \quad {}+\frac{\b}{n-2}\left[\frac{p-1}{n+1-2p}\frac{\E^2}{r^{2(n-p-1)}}-\frac{1}{p(n-1-2p)}\frac{\B^2}{r^{2(p-1)}}\right] \quad (2p\neq n\pm1), \label{H_generic}
\end{eqnarray}
where $\Lambda$, $\mu$, $\E^2$, $\B^2$ and $K=0,\pm 1$ are constants. There is a Killing vector field $\pa_u$ (so that the metric is static in regions where $H>0$) and roots of $H(r)$ define Killing horizons (see also \cite{BarCalCha12}). In the special cases $2p=n\pm 1$ ($n$ odd), the second line of \eqref{H_generic} should be replaced by appropriate logarithimc terms (which in fact vanish for $n=5,7$, see \cite{OrtPodZof15} for details). 

The corresponding ``Coulombian'' Maxwell field is given by 
\be
 \bF=\frac{1}{(p-2)!}\frac{1}{r^{n+2-2p}}\e_{i_1\ldots i_{p-2}}(x)\,\d u\wedge\d r\wedge\d x^{i_1}\wedge\ldots\wedge\d x^{i_{p-2}}+\frac{1}{p!}\bb_{i_1\ldots i_{p}}(x)\,\d x^{i_1}\wedge\ldots\wedge\d x^{i_{p}} , \label{F_generic}
\ee
where $\e_{i_1\ldots i_{p-2}}$ and $\bb_{i_1\ldots i_{p}}$ are harmonic forms (of respective rank $(p-2)$ and $p$) in the transverse geometry $h_{ij}$. These forms are further constrained by 
\beqn
 & & \E^2_{ij}=\frac{\E^2}{n-2}h_{ij} \quad {(p\ge 3)} , \qquad \B^2_{ij}=\frac{\B^2}{n-2}h_{ij} \quad {(p\le n-2)} , \label{EB_generic} \\
 & \mbox{where } \ \ \  & \E_{ij}^2\equiv \e_{ik_1\ldots k_{p-3}}\,\e_{j}^{\ k_1\ldots k_{p-3}}  , \qquad \E^2\equiv h^{ij}\,\E_{ij}^2 \qquad (p\ge 3) , \label{E} \\
 & & \B_{ij}^2\equiv \bb_{ik_1\ldots k_{p-1}}\,\bb_{j}^{\ k_1\ldots k_{p-1}} , \qquad \B^2\equiv h^{ij}\,\B_{ij}^2 \qquad (p\le n-2) . \label{B}
\eeqn
Conditions \eqref{EB_generic} also impose severe restrictions on the Einstein metric $h_{ij}$ \cite{OrtPodZof08,BarCalCha12,OrtPodZof15}.

The above solutions are extensions to arbitrary $p$ of the $p=2$ ($n\neq6$) solutions studied in \cite{OrtPodZof08} (including, when $\B^2=0$, the $n>4$ Reissner-Nordstr\"{o}m solution found by Tangherlini \cite{Tangherlini63}) and possess similar properties.  They were previously obtained in \cite{BarCalCha12} (using a different method) and represent static black holes dressed with electric and magnetic Maxwell fields, at least for certain values of the parameters in \eqref{H_generic}. However, the metric $h_{ij}$ in \eqref{ds_generic} cannot describe a round sphere\cite{OrtPodZof15}, except when $p=2$ and $\bb_{i_1 i_{2}}=0$ (or, dually, when $p=n-2$ and $\e_{i_1\ldots i_{n-4}}=0$), so that these black holes cannot have a spherical horizon and cannot be asymptotically flat (in agreement with  \cite{Guven89,EmpOhaShi10,ShiOhaTan11}), except in the electric $p=2$ (or magnetic $p=n-2$) Reissner-Nordstr\"{o}m solution of \cite{Tangherlini63}. A flat and compact horizon metric $h_{ij}$ is instead permitted (giving $K=0$ in \eqref{H_generic}; then the harmonic forms $\e_{i_1\ldots i_{p-2}}$ and $\bb_{i_1\ldots i_{p}}$ must be covariantly constant \cite{Bochner48,YanBocbook}), thus allowing, e.g., for asymptotically locally (A)dS black holes (see also \cite{BarCalCha12}). An additional ``Vaidya-type'' matter field representing pure radiation aligned with $\bk$ can easily be included by appropriate simple modifications \cite{OrtPodZof08,BarCalCha12,OrtPodZof15}.

The Weyl type of the spacetimes is D(bd) and $\bk=\pa_r$, $\bl={-}\pa_u+H\pa_r$ are the (unique) double WANDs. These are manifestly also aligned null directions of the Maxwell field \eqref{F_generic}, which is thus also of type D. Also the Ricci tensor is of (aligned) type D. See \cite{Tangherlini63,OrtPodZof08,BarCalCha12,OrtPodZof15} for explicit examples with various values of $n$ and $p$.  

Let us just observe that more general solutions are permitted in the {\em exceptional cases $2p=n\pm 2$ ($n\ge6$, even)}, which are in general non-static and of Weyl type II(bd) -- see \cite{OrtPodZof15} for details.

\section{Special case $2p=n$ ($n$ even): black holes with electromagnetic radiation}

\label{subsec_2p=n_summary}

Static black hole solutions clearly exist also for the special rank $2p=n$. However, there exists now also a new family of solutions which allows for $F_{u j_1 \ldots j_{p-1}}\neq 0$. The Maxwell field is given by
\beqn
 \bF= & & \frac{1}{\left(\frac{n}{2}-2\right)!}\frac{1}{r^{2}}\e_{i_1\ldots i_{p-2}}\d u\wedge\d r\wedge\d x^{i_1}\wedge\ldots\wedge\d x^{i_{p-2}}+\frac{1}{\left(\frac{n}{2}\right)!}\bb_{i_1\ldots i_{p}}\d x^{i_1}\wedge\ldots\wedge\d x^{i_{p}} \nonumber \\
			& & {}+\frac{1}{2\left(\frac{n}{2}-1\right)!}\left(-\frac{n-2}{r}\e_{[i_2\ldots i_{p-1},i_1]}+2{f}_{i_1\ldots i_{p-1}}\right)\d u\wedge\d x^{i_1}\wedge\ldots\wedge\d x^{i_{p-1}} . \label{F_2p=n}
\eeqn
The forms $\e_{i_1\ldots i_{p-2}}(u,x)$ and $\bb_{i_1\ldots i_{p}}(u,x)$ are generally not harmonic in the transverse space, but satisfy the ``modified'' Euclidean Maxwell equations in $(n-2)$ dimensions 
\beqn
	& & (\sqrt{h}\,\e_{}^{ji_1\ldots i_{p-3}})_{,j}=0 \quad (p\ge 3) , \qquad {\bb_{[i_1\ldots i_p,j]}= 0 } , \label{diverg_elec} \\
	& & {(\sqrt{h}\, \bb^{kj_1 \ldots j_{p-1}})_{,k}}= \frac{1}{2}(n-2)\sqrt{h}\, h^{i_1 j_1} \ldots h^{i_{p-1} j_{p-1}} \e_{[i_2 \ldots i_{p-1}, i_1]} . \label{maxw_2p_1} 
\eeqn
In addition, they can depend on $u$ via
\be
	\bb_{i_1 \ldots i_{p},u}=\frac{n}{2}{f}_{[i_2 \ldots i_{p},i_1]} , \qquad {(\sqrt{h}\, \e^{i_1 \ldots i_{p-2}})_{,u} = (\sqrt{h}\, {f}^{k i_1 \ldots i_{p-2}})_{,k} .} \label{maxw_2p_2}
\ee
The latter further tells us that the $(p-1)$-form ${f}_{i_1\ldots i_{p-1}}(u,x)$ is also generically non-harmonic.

The line-element is given by \eqref{ds_generic} with
\be 2H=K{+}\frac{2(\ln\sqrt{h})_{,u}}{n-2}\,r-\frac{2\Lambda}{(n-1)(n-2)}\,r^2-\frac{\mu}{r^{n-3}} +\frac{\b}{2}\left(\E^2+\frac{4}{n(n-2)}\B^2\right)\frac{1}{r^{n-2}} , \label{grr_2p=n}
\ee
where $\E^2$, $\B^2$ and $\mu$ are generically functions of $u$ and $x$, cf. the corresponding equations given in \cite{OrtPodZof15}. The metric $h_{ij}(u,x)=h^{1/(n-2)}(u,x)\,\gamma_{ij}(x)$ is again Einstein ($\gamma_{ij}(x)$ is unimodular), with scalar curvature $\R=K(n-2)(n-3)$ (where $K=0,\pm 1$), with the constraint (recall \eqref{E}, \eqref{B})
\be
 \frac{1}{4}(n-2)(n-4)\left(\E^2_{ij}-\frac{\E^2}{n-2}h_{ij}\right)=\B^2_{ij}-\frac{\B^2}{n-2}h_{ij} . \label{eqEB}
\ee

Considerable simplifications arise if $h_{ij}$ is taken to be the metric of a compact space\cite{OrtPodZof15}. In particular, {\em $h_{ij}$ cannot describe a round sphere}, and no asymptotically flat spacetimes are thus to be found in this class of solutions. 
In general, the Weyl tensor is of type II(bd). The type D(bd) (or D(bcd)) is possible in special cases, but the types III, N and O are forbidden when $\bF\neq0$. 

The Maxwell field \eqref{F_2p=n} is in general of type II (aligned with $\bk$ by construction) and, in a parallelly transported frame adapted to $\bk$, peels off as $\bF=\frac{\mbox{\boldmath{$N$}}}{r^{\frac{n}{2}-1}}+\frac{\mbox{\boldmath{$II$}}}{r^{\frac{n}{2}}}$ (in agreement with test-field results \cite{Ortaggio14}). It becomes of type N iff $\e_{i_1\ldots i_{p-2}}=0=\bb_{i_1\ldots i_{p}}$, and of type D under the conditions given in\cite{OrtPodZof15}. When $p$ is odd, a self-dual $\bF$ is possible in special cases. 

For $n=6=2p$, a simple example is given by
\beqn
 & & h_{ij}=\delta_{ij} , \qquad 2H=-\frac{\Lambda}{10}r^2-\frac{\mu(u)}{r^3} , \qquad \mu(u)=\mu_0{-}\frac{\b}{2}\int\F^2\d u , \nonumber \\
 & & F_{uij}=f_{ij}(u) . \label{6D_null}
\eeqn
For $\Lambda<0$ this spacetime represents the formation of asymptotically locally AdS black holes with electromagnetic radiation. By a rotation, one can always simplify the Maxwell field so as to only have non-zero components $F_{u12}=f_{12}(u)$, $F_{u34}=f_{34}(u)$, in which case $\bF$ is (anti-)self-dual when $f_{34}(u)=\mp f_{12}(u)$. Solution~\eqref{6D_null} is an extension of a solution given for $n=4$ (for $\Lambda=0$) in \cite{RobTra62} and recently discussed in \cite{Senovilla15}. 

See \cite{OrtPodZof15} for more details and other examples.

\section*{Acknowledgments}


This work has been supported by the Albert Einstein Center for Gravitation and Astrophysics, Czech Science Foundation GA\v{C}R~14-37086G. M.O. has also been supported by research plan {RVO: 67985840}.

%
%
%


\end{document}